\documentclass{aastex62}

\shorttitle{CO in Hot Jupiter Around CI Tau}
\shortauthors{Flagg et al.}

\begin{document}


\title{CO Detected in CI Tau b: Hot Start Implied by Planet Mass and M$_K$}


\author{Laura Flagg}
\affil{Department of Physics and Astronomy, Rice University, 6100 Main St. MS-108, Houston, TX 77005, USA}
\email{laura.flagg@rice.edu}

\author{Christopher M. Johns-Krull}
\affil{Department of Physics and Astronomy, Rice University, 6100 Main St. MS-108, Houston, TX 77005, USA}
\affil{Visiting Astronomer at McDonald Observatory of The University of Texas at Austin}

\author{Larissa Nofi}
\affil{Lowell Observatory, 1400 W Mars Hill Road, Flagstaff, AZ 86001, USA}
\affil{Institute for Astronomy, University of Hawai’i Manoa, 2680 Woodlawn Drive, Honolulu, HI 96822, USA}

\author{Joe Llama}
\affil{Lowell Observatory, 1400 W Mars Hill Road, Flagstaff, AZ 86001, USA}

\author{L. Prato}
\affil{Visiting Astronomer at McDonald Observatory of The University of Texas at Austin}
\affil{Lowell Observatory, 1400 W Mars Hill Road, Flagstaff, AZ 86001, USA}

\author{Kendall Sullivan}
\affil{McDonald Observatory and the Department of Astronomy, The University of Texas at Austin, Austin, TX 78712, USA}

\author{D. T. Jaffe}
\affil{McDonald Observatory and the Department of Astronomy, The University of Texas at Austin, Austin, TX 78712, USA}

\author{Gregory Mace}
\affil{McDonald Observatory and the Department of Astronomy, The University of Texas at Austin, Austin, TX 78712, USA}

%
%
%
%

\begin{abstract}
We acquired high resolution IR spectra of CI Tau, the host star of one of the few young planet candidates amenable to direct spectroscopic detection.  We confirm the planet's existence with a direct detection of CO in the planet's atmosphere.  We also calculate a mass of 11.6 M$_J$ based on the amplitude of its radial velocity variations.  We estimate its flux contrast with its host star to get an absolute magnitude estimate for the planet of 8.17 in the K band.  This magnitude implies the planet formed via a ``hot start'' formation mechanism. This makes CI Tau b the youngest confirmed exoplanet as well as the first exoplanet around a T Tauri star with a directly determined, model-independent, dynamical mass. 
\end{abstract}





\section{Introduction}
Observations of young exoplanets are key to understanding planet formation. Models for giant planets indicate planets can form with very diﬀerent temperatures at early ages. It  is  often  assumed  that  gravitational  instability produces hot planets
while core accretion forms comparatively cooler and less luminous planets \citep{MarleyLuminosityYoungJupiters2007}, although it may be possible for both formation models to produce planets at a range of temperatures \citep{HelledGiantPlanetFormation2014}.  To better distinguish between these possibilities, young planets with model-independent masses are needed.  Several planets have been found via direct imaging around pre-main sequence stars, including some around T Tauri stars \citep[e.g.,][]{Chauvingiantplanetcandidate2004, IrelandTwoWidePlanetarymass2011, RameauDiscoveryProbable452013}, and their measured magnitudes have been used to evaluate planet formation theories \citep{MarleauConstraininginitialentropy2014, RajanCharacterizing51Eri2017}.  However, for these systems, the planet mass estimates are necessarily model-dependent as it is generally not possible to detect the stellar reflex motion given the extreme mass ratios and the large star-companion separations.  With the exception of $\beta$ Pic b \citep{Snellenmassyoungplanet2018, DupuyModelindependentMassModerate2019}, confirmation of these objects as planetary mass companions is not yet possible.


Young stars are significantly more active than main sequence stars, which hides the subtle planetary signals, making detection via either the radial velocity (RV) or the transit method difficult.    However, in the past several years, planet candidates have been reported around pre-main sequences stars using the RV method including CI Tau b \citep{Johns-KrullCandidateYoungMassive2016}, V830 Tau b  \citep{DonatihotJupiterorbiting2016}, and Tap 26 b \citep{YuhotJupitervery2017}.    These targets would be ideal for evaluating planetary evolution models if the planet's brightness or temperature could be determined. To do so, the planet needs to be detectable directly.  By virtue of being close enough to their host stars to induce substantial gravitational reflex motion, these planets are too close to be detected via direct imaging.

Direct spectroscopic detection  has been used to measure both planet masses and contrast ratios between absorption lines in hot Jupiter atmospheres and their main sequence host stars' continuum \citep[e.g.][]{Snellenorbitalmotionabsolute2010, Brogisignatureorbitalmotion2012, BirkbyDetectionwaterabsorption2013, RodlerDetectionCOabsorption2013}.  Near infrared spectra are used to detect molecules common in planetary atmospheres like H$_2$O and CO.   By using spectra from different phases in the planet's orbit, the median stellar signal can be simply subtracted out, while the planet's signal, which is moving at hundreds of kilometers per second, remains, allowing for a direct detection of the planet's spectrum.

CI Tau is a $\sim$2-3 Myr old, K7, classical T Tauri star.  \citet{Johns-KrullCandidateYoungMassive2016} detected a M$_P$sin\textit{i}=8.08$\pm$1.53  M$_J$ planet candidate around it with a 8.989 day period based on analysis of its RV variations.  Based on an inclination of $\sim44^\circ$ estimated from imaging its disk  \citep{Guilloteaumassesyoungstars2014}, its total mass is $\sim$11 M$_J$.  If the planet formed via some ``hot start'' mechanism, its expected contrast ratio with its host star should be a few$\times$10$^{-3}$ based on models from \citet{SpiegelSpectralPhotometricDiagnostics2012} and thus would be directly detectable spectroscopically.  If, however, the planet formed via a ``cold start'' method, its contrast ratio would likely be $\sim10^{-5}$ at best, a ratio no current techniques are sensitive to.


\section{Observations and Data Analysis}
\subsection{Observations}
Our 40 observations of CI Tau were taken between 2014-2018 with the Immersion Grating Infrared Spectrograph (IGRINS) at both the McDonald Observatory 2.7-m Harlan J. Smith Telescope and the Lowell Observatory 4.3-m Discovery Channel Telescope. CI Tau was observed by nodding between the A and B positions on the slit.  Total exposure times ranged from 240 to 2400 seconds, resulting in final S/N values of 125-300. Each target observation was paired with a standard A0V star observation at a similar airmass used for telluric corrections. IGRINS makes use of a silicon immersion grating and a fixed optical path to simultaneously cover the H and K bands (1.45-2.5$\mu$m) with a spectral resolving power R$\sim$45,000. Additional discussion on the design and capabilities of IGRINS can be found in \citet{ParkDesignearlyperformance2014,  MaceIGRINSDiscoveryChannel2018}. 

\subsection{Data Reduction}
We reduced the data with the IGRINS pipeline package \citep[version 2.2.0 alpha 1;][]{jae_joon_lee_2017_845059}. The pipeline implements flat-fielding, bad pixel correction, sky subtraction, and source extraction based on the methods of \citet{Horneoptimalextractionalgorithm1986}. The pipeline wavelength solution is based on telluric emission and absorption lines.  



To continuum normalize, we fit a 6th order polynomial to each order of the spectrum of an A0V star taken the same night.  We scale this fit to account for differences in the maximum flux, and then divide each CI Tau spectrum by the scaled fit to remove the blaze function. Because it is critical to have all observations on the same wavelength scale, we then cross-correlate the 2.06 - 2.08 $\mu m$ order with a telluric template.  At these wavelengths, the spectrum is dominated by telluric features.  We corrected for any discrepancies by shifting the wavelength solution for the all CI Tau spectral orders by the offset between the telluric lines in the target and in the A0V spectrum. All spectra are then interpolated onto a common wavelength scale with a spacing of 2.04 km/s per pixel to match the native resolution of IGRINS.

\subsection{Removing Telluric and Stellar Features}



We used spectra of A0V stars to correct for the telluric features.  As the telluric spectra were generally not taken at the exact same airmass as the target, we raised the flux of the continuum normalized telluric spectrum to a power to account for changes in line strength.  This power was determined by fitting each A0V spectra using the Levenberg-Marquardt algorithm to perform a least-squares minimization \citep{NewvilleLMFITNonLinearLeastSquare2014} until the telluric lines in the A0V star matched the depth of the telluric lines in the target spectra.  We do not currently adjust for compositional differences in the telluric spectra, so we use A0V  spectra taken close in time to the target spectra. Our method for removing telluric features worked relatively well in regions where there is not too much telluric absorption.  However, there are regions of the spectrum that correct poorly with this method because of very strong telluric absorption.  We chose to mask those regions out of the analysis.  After telluric correction, we then corrected each CI Tau spectrum for barycentric velocity differences.

 CI Tau has a protoplanetary disk, therefore we also need to correct for differences in the CI Tau spectra due to continuum veiling, the ratio of the excess disk flux to the flux from the stellar continuum.  We took the observation with the least amount of veiling and de-veil all other observations to match that one.    We estimated the overall veiling of this least veiled CI Tau spectrum by fitting it to a template spectrum accounting for differences in rotational broadening, following \citet{Johns-KrullNewInfraredVeiling2001}.    For the template, we  use IGRINS spectra of V830 Tau, a diskless star of the same age and spectral type.  While \citet{DonatihotJupiterorbiting2016} report that V830 Tau has a planet, the strength of any lines contributed by the planet is expected to make a negligible change in the derived veiling.  
 
 After correcting each observation for telluric absorption and veiling, the stellar spectrum ideally should not change except for the signal of the planet moving in velocity space. To ensure we have removed as much of the star's signal as possible, we needed to correct for any slight differences in velocity. We cross-correlated each stellar spectrum with the highest signal to noise spectrum, so that we could Doppler shift them to the same stellar frame.  We then subtracted the median spectrum, ideally leaving only the planet's signal, and shifted the spectra back to their original frame. T Tauri stars' variability arises from other processes in addition to veiling.  As a result, it is difficult to account for all variability  in the target spectra even after subtracting the median (see section \ref{res}).
 
 
 \subsection{Planet Detection via Cross-Correlation}\label{methods:scx}
We employ two methods to magnify the weak signal from the planet.  One is by phase shifting the spectra so the planet's signal from observations at different phases of the planet's orbit, initially at different of RVs, is aligned;  we then co-added all the spectra. The other is cross-correlating the spectrum with a template which combines the signal from all the spectral features into one signal. We chose to base our analysis by first co-adding and then cross-correlating; however, reversing the order
also results in a detection as described below.


 
 In order to shift the spectra to account for the planet's orbital motion, its orbit must already be known.  We fixed the orbital parameters at the values listed in Table \ref{params} from \citet{Johns-KrullCandidateYoungMassive2016}.  However, the stellar inclination is not derived from the RV curve. The velocity amplitude of the star could be caused by a more massive planet on a more face-on orbit or a less massive planet on a more edge-on orbit.  This also impacts the RV shift of the planet with a larger planetary velocity amplitude, K$_{P}$, occurring if the orbit is more edge on. As we cannot assume a K$_{P}$ from the stellar RV curve, we vary K$_{P}$ in increments of 1 km/s over the region of physically allowable parameter space to look for the strongest signal.  For each value of K$_{P}$, we calculate the cross-correlation function at a range of offsets between -400 and 400 km/s in increments corresponding to the velocity spacing of the pixels.  The template for the cross-correlation is created from VALD \citep{RyabchikovamajorupgradeVALD2015}, which used the CO line list from \citet{KCO},  for an object of T$_{eff}$=2500 K with log\textit{g}=4.0 dex.   Because of the reliance on the CO lines for the reported detection, we used only 4 orders of the spectra that covered wavelengths between 2.29 and 2.42 $\mu m$.


\section{Results}\label{res}
\begin{figure}[!ht]
\centering
\includegraphics[width=6.5in]{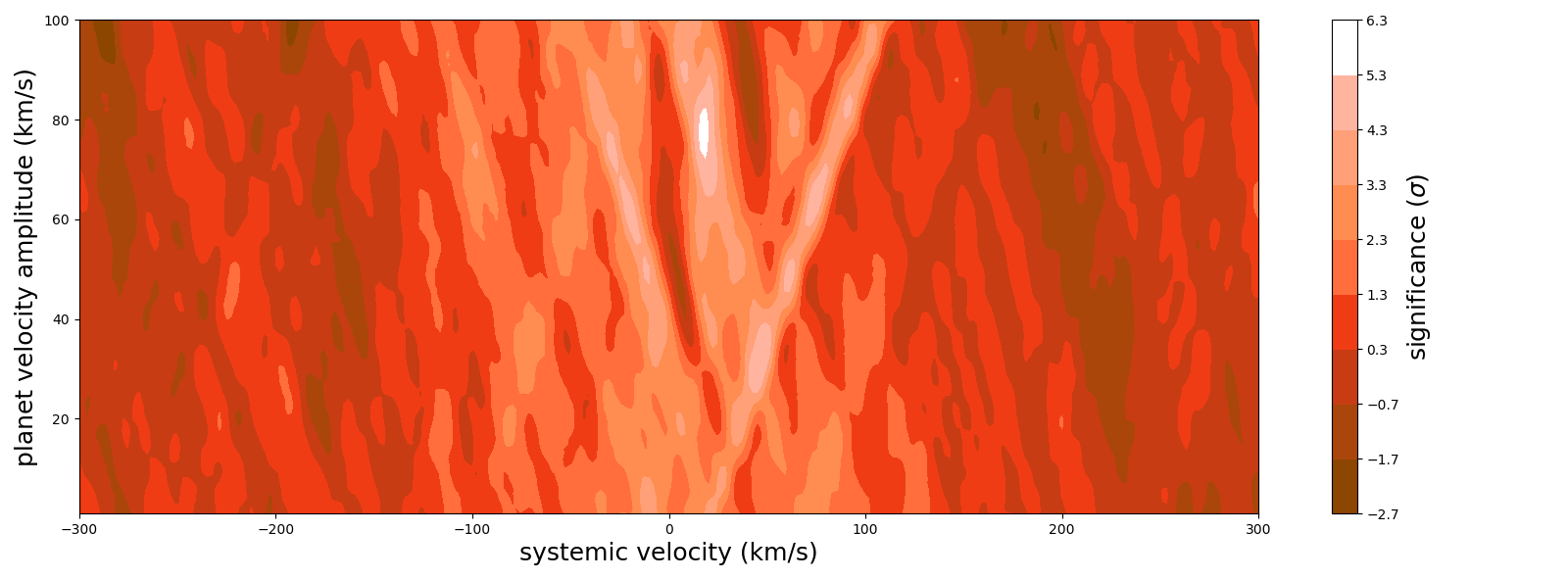}
\caption{The cross-correlation function significance plotted as a function of planet velocity amplitude and systemic radial velocity.  The significance is the number of standard deviations above the mean for those parameters, peaking at 5.7$\sigma$.
 \label{velocity}}
\end{figure}

The cross-correlation matrix (Figure \ref{velocity}) peaks at at a systemic RV, V$_{sys}$=18.0$\pm_{4.4}^{3.7}$ km/s and K$_{P}$=77.4$\pm_{9.4}^{7.0}$ km/s.  The value of V$_{sys}$ is consistent within uncertainties to that measured by \citet{TorresSearchassociationscontaining2006}. To estimate the significance of this detection, we adapted the method from \citet{BirkbyDetectionwaterabsorption2013}.  We took the CCF matrix, subtracted the mean CCF value and divided by the standard deviation of the values away from the signal as in \citet{HoeijmakersAtomicirontitanium2018}.  Using this, the peak has a significance of 5.7$\sigma$. If we first cross-correlate each individual CI Tau spectrum with the template and then sum the cross correlation functions, the peak occurs at V$_{sys}$ = 17.1 km/s and K$_{P}$ = 77.9 km/s (both equal to within the uncertainties to the values above) with a significance of 4.8$\sigma$. We also detect the signal in each of the four orders and each observing season separately.  The 1$\sigma$ uncertainty in K$_{P}$ and V$_{sys}$  is taken as the points in the CCF matrix where the CCF value has dropped by 1, i.e from 5.7 to 4.7, as in \citet{BrogiExoplanetatmospheresGIANO2018}.  To verify this  uncertainty, we have created a bootstrap sample \citep{EfronBootstrapMethodsAnother1979} of N=10000 data sets of 40 spectra, randomly sampling with replacement from our original 40 spectra.  We then calculate V$_{sys}$ and K$_P$ for each of the N data sets, creating a new distribution with N values for each parameter. This resulted in measurements of V$_{sys}$=18.3$\pm_{1.8}^{2.7}$ km/s and K$_{P}$=76.9$\pm_{5.6}^{4.9}$ km/s, with the 1$\sigma$ error bars corresponding to the 68.3\% confidence interval from these distributions. We chose to be conservative and use the larger uncertainties above for further analysis.  Uncertainties in K$_P$ and K$_*$ dominate the uncertainties in the additional quantities derived below. 



 \begin{deluxetable}{lrr}
 \tabletypesize{\normalsize}
 \tablewidth{0pt}
 \tablecaption{Parameters of the CI Tau System}
 \tablehead{
 \colhead{Parameter} & \colhead{    \hspace{.25cm}   } & \colhead{Value}
  }
 \startdata
M$_*$ (M$_\odot$) & & 0.90$\pm$0.02 \\
K$_*$ (km/s) & & 0.950$\pm$0.207 \\
distance (pc) & & 158.0$\pm$1.2 \\
   \hline
orbital period (days) & & 8.9891$\pm$0.0202  \\
eccentricity  & & 0.25$\pm$0.16 \\
angle of periastron ($^\circ$) &  & 31$\pm$56 \\
\hline
K$_P$ (km/s) & & 77.4$\pm_{9.4}^{7.0}$ \\
M$_P$ (M$_J$) & & 11.6$\pm_{2.7}^{2.9}$\\
\enddata
\tablecomments{Distance from Simon et al. (in prep), based on Gaia \citep{BrownGaiaDataRelease2018} and \citet{Bailer-JonesEstimatingDistanceParallaxes2018}.  Stellar mass,  initially measured by \citet{Guilloteaumassesyoungstars2014} adjusted for distance and the planet's mass.  \label{params} }
 \end{deluxetable}

Based on the determined value of K$_{P}$ and the  velocity amplitude of the star, K$_{*}$, we calculated a mass ratio $\frac{M_*}{M_P}$ of 81.5$\pm_{20.3}^{19.2}$.   We assume a stellar mass of 0.9 M$_\odot$.   This leads to a planet mass of 11.6$\pm_{2.7}^{2.9}$ M$_J$.     Our detection is fully consistent with planet parameters reported in \citet{Johns-KrullCandidateYoungMassive2016}.   Based on the mass of the planet and its separation with its host star, models indicate that it could not form at its current location  \citep[and sources therein]{DawsonOriginsHotJupiters2018}.  Thus, the planet likely formed elsewhere in the disk and then moved inward.


We have calculated an inclination of 50.5$\pm_{8.5}^{6.3}$ degrees, using sin(i)=$\frac{K_P}{V_P}$, where V$_P$ is the orbital velocity amplitude calculated from Kepler's Laws using the assumed stellar mass, eccentricity, and period.  This is within uncertainties of disk inclinations previously measured, which are in the range of 44 to 51$^\circ$ \citep{Guilloteaumassesyoungstars2014, TripathimillimeterContinuumSizeLuminosity2017, ClarkeHighresolutionMillimeterImaging2018}. Given these inclinations, we do not expect the planet to transit.

The cross-correlation plot (Figure \ref{velocity})  shows residuals that we believe come from the star itself.  Unlike host stars that are typical targets for direct detection, CI Tau is a relatively cool star and thus has some of the same CO features we are attempting to detect in the planet.  Given the star is young and variable, our method likely does not completely remove the star's signal from the spectra.   For example, the spot coverage on the star is expected to vary which should produce small differences in the CO spectra coming from the star.  As a result, there are likely artifacts from the star that show up in Figure \ref{velocity}, specifically the v-shaped feature that starts from the lower axis of the plot at the systemic velocity of the star, which is likely due to CI Tau itself. 

Of particular interest from this detection is the contrast ratio between the star and the planet.  By inserting simulated planets at random systemic and orbital velocities, we can estimate the contrast ratio needed to produce the signal we see.  Our best estimate for the flux ratio between the star+disk and planet is 356$\pm_{69}^{94}$. With our estimated veiling of 1.08$\pm$0.16 for the least veiled observation of  CI Tau, (which is consistent with other veiling measurements for CI Tau - Sokal et al., submitted) this results in a star to planet flux ratio of 172$\pm_{37}^{54}$.  Based on the contrast ratio between the planet and the star+disk,  CI Tau's apparent magnitude of 7.793 in the K band from 2MASS \citep{SkrutskieTwoMicronAll2006}, and the system's distance, we calculate an absolute magnitude of 8.17$\pm_{0.21}^{0.33}$ in the K band for CI Tau b.

\section{Discussion and Conclusions}
We have analyzed K band spectra of CI Tau and found strong support of the existence of CI Tau b, a hot Jupiter with a mass of 11.6$\pm_{2.7}^{2.9}$ M$_J$. We have directly detected CO in CI Tau b's atmosphere. We have also estimated its magnitude in the K band.   These results make CI Tau b the youngest confirmed exoplanet and the only exoplanet around a T Tauri star with a model-independent mass. 

\begin{figure}[!ht]
\centering
\includegraphics[width=3.4in]{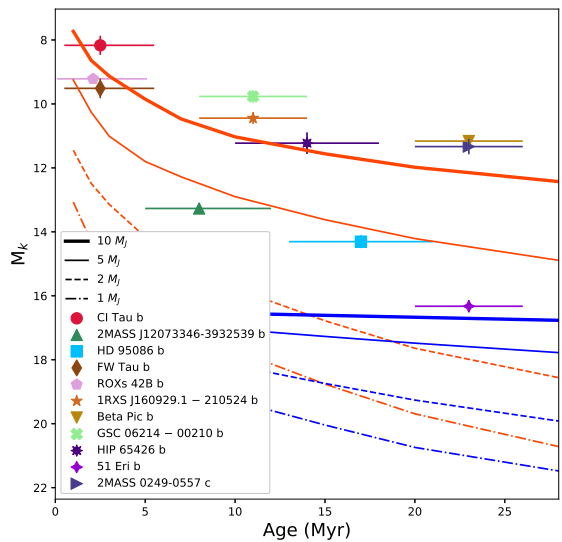}
\caption{Absolute K magnitude versus age for a selection of planetary mass companions.  In red are the hot start tracks from \citet{SpiegelSpectralPhotometricDiagnostics2012}, while the cold start tracks are in blue.  The known planetary mass companions around the youngest stars appear consistent with a hot start formation.  Magnitudes for 2MASS J12073346-3932539, HD 95086 b, 1RXS J160929.1 − 210524 b, $\beta$ Pic b, HIP 65426 b, and Eri 51 b
 are from \citet{Chauvingiantplanetcandidate2004, DeRosaSpectroscopicCharacterizationHD2016, LafreniereDirectImagingSpectroscopy2008, BonnefoyHighangularresolution2011, IrelandTwoWidePlanetarymass2011, ChauvinDiscoverywarmdusty2017, RajanCharacterizing51Eri2017}; and \citet{DupuyHawaiiInfraredParallax2018} respectively.  Magnitudes for FW Tau b and ROXs 42B b are from \citet{KrausThreeWidePlanetarymass2014}. 
 \label{youngplanets}}
\end{figure}

Based on models of young, giant exoplanets \citep{SpiegelSpectralPhotometricDiagnostics2012}, a planet would need to form via a ``hot start'' mechanism to be this bright at a 10 M$_J$ planetary mass.  This confirms hot start models as a viable scenario for planet formation for the case of CI Tau b. In addition, according to \citet{SpiegelSpectralPhotometricDiagnostics2012}, a 10 M$_J$ planet at 2 Myr would have a radius of $\sim$2 R$_J$ and a T$_{eff}\sim$2300 K; \citet{BaraffeNewevolutionarymodels2015} give similar numbers for a 0.01 M$_\odot$ (10.5 M$_J$) object at that age.   This bright magnitude is consistent with other planetary mass companions around young stars, which also appear consistent with a hot start formation (Figure \ref{youngplanets}).  However, other than $\beta$ Pic b and now CI Tau b, thus far none of those objects have model independent masses needed to confirm them as planets. 

Even a cold-start planet can have a high surface temperature if its surface is externally heated.  Assuming the planet  is tidally locked, we can estimate its equilibrium temperature as $T_{eq}=\left(\frac{(1-A_b)L_S}{8\pi \sigma d^2}\right)^{1/4}$, with L$_*$  the host star's total luminosity, $d$ the separation, $\sigma$ the Stefan-Boltzman constant, and $A_b$ the planet's albedo \citep{MarleyAtmospheresExtrasolarGiant2007}. We combined the intrinsic stellar luminosity of 0.6 L$_\odot$ \citep{McClureCharacterizingStellarPhotospheres2013} with the accretion luminosity of 0.3 L$_*$ \citep{ValentiTauriStarsBlue1993}, for a total source luminosity of 0.78 L$_\odot$.  Based on this, we calculated temperatures ranging from 820 to 1080 K, depending on the planet's albedo.  There also may be heating from the inner disk wall. \citet{McClureCharacterizingStellarPhotospheres2013} gives an inner wall luminosity of 0.08 L$_*$.  We estimated the flux from the disk that reaches the planet when at apastron by integrating the flux from the inner disk wall over its area, using the height and radius of the disk wall from \citet{McClureCurvedWallsGrain2013}.   Adding this flux to the flux from the star and accretion luminosity results in temperatures between 1000 and 1300 K.  Thus, it is unlikely that CI Tau b's proximity to its host star or the disk is responsible for its bright K band magnitude.  







While historically, hot starts have been associated with gravitational instability, some recent research indicates core accretion may be able to produce hot start planets \citep[e.g.][]{BerardoHotstartGiantPlanets2017}.  As a result, more observations are needed. Detecting other hot Jupiters in a similar manner over a range of ages would help put additional limits on formation models. We expect more planets to be found and confirmed around very young stars with high precision RV analysis.  Such planets will be excellent targets for analysis similar to that presented here.   Composition could also be indicative of formation method \citep{ObergEffectsSnowlinesPlanetary2011}. Additionally, our understanding of the closest regions to a young star is still very limited. Further work is needed to better understand the environment the planet is in, including on the impact of the inner disk wall on young hot Jupiters. 

We will extend this research in the future. While the planet is well detected despite fixing most orbital parameters, this may result in underestimated uncertainties for K$_P$ and other derived quantities.  We plan to employ MCMC to explore the sensitivity of our results to the assumed orbital parameters and how the quantities we measure, and their uncertainties, are affected.  This should result in more accurate measurements of the planet mass and magnitude. Second, using the data we currently have between 1.4 and 2.5 $\mu$m, we will search for additional molecules.  We are currently only using four out of over 40 orders of spectroscopic data, and while not all of the orders are useful, if H$_2$O is present, it could be detected within these data as well.   This could inform the C/O ratio and put additional limitations on how the planet formed \citep{ObergEffectsSnowlinesPlanetary2011}.  In addition, by using model atmosphere templates at different temperatures, in future work we will estimate the planet's temperature directly.  

\section*{Acknowledgements}
We thank the referee for their helpful comments. This research was supported in part by a Faculty Initiative Fund grant from Rice University.  CMJ-K would like to acknowledge partial support from the NSF through an INSPIRE grant to Rice University (grant AST-1461918), as well as partial support from NASA through grant 80NSSC18K0828.   IGRINS was developed with the financial support of the US National Science Foundation under grants AST-1229522 and AST-1702267, the University of Texas at Austin, and the Korean GMT Project of KASI. This work has made use of the VALD database, operated at Uppsala University, the Institute of Astronomy RAS in Moscow, and the University of Vienna.  This work has made use of data from the European Space Agency (ESA) mission {\it Gaia} (\url{https://www.cosmos.esa.int/gaia}), processed by the {\it Gaia} Data Processing and Analysis Consortium (DPAC, \url{https://www.cosmos.esa.int/web/gaia/dpac/consortium}). Funding for the DPAC has been provided by national institutions, in particular the institutions participating in the {\it Gaia} Multilateral Agreement. This research has made use of the VizieR catalogue and SIMBAD database, operated at CDS, Strasbourg, France  \citep{WengerSIMBADastronomicaldatabase2000}.  This work made use of PyAstronomy (\url{https://github.com/sczesla/PyAstronomy}).

\bibliography{citaub}
\end{document}